\documentclass[%
 preprint, linenumbers,
 amsmath,amssymb,
 aps, physrev,
floatfix,
]{revtex4-2}

\usepackage{graphicx}
\usepackage{dcolumn}
\usepackage{bm}
\usepackage{float}
\usepackage{makecell}
\usepackage{pifont}
\usepackage{url}
\usepackage{hyperref}
\usepackage{cleveref}
\begin{document}
\nolinenumbers

\preprint{APS/123-QED}

\title{\textbf{Interaction of a Vortex Pair with a Polymeric Fluid Layer} 
}%

\author{Rabia Sonmez}
 \email{Contact author: rsonmez@gmu.edu}
\affiliation{%
Department of Mechanical Engineering\\
George Mason University, Fairfax, VA 22030, USA
}%

\author{Robert A. Handler}
\affiliation{%
Center for Simulation and Modeling \\
George Mason University, Fairfax, Virginia 22030, USA
}%
\author{David B. Goldstein}
\author{Anton Burstev}
\author{Ryan Kelly}
\affiliation{
 Department of Aerospace Engineering\\ 
 University of Texas, Austin, TX 2617, USA
}%
\author{Saikishan Suryanarayanan}
\affiliation{%
 Department of Mechanical Engineering\\
 University of Akron, Akron, OH 44325, USA}%

\date{\today}

\begin{abstract}
\noindent The interaction of vortical structures with boundaries has been extensively studied in Newtonian fluids, where conditions such as no-slip walls, free surfaces, or contaminated surfaces dictate whether vortices rebound, dissipate, or generate secondary structures. In this work, we investigate a related but fundamentally different problem: the interaction of a vortex pair with a finite, non-uniform layer of polymeric fluid. Numerical simulations employing the finitely extensible nonlinear elastic-Peterlin model  are used to examine the effects of polymer concentration, relaxation time, polymer layer thickness, and maximum polymer extension on the evolution of kinetic energy and enstrophy. The results show that, while the polymeric fluid dissipates vortical motion, vortex-polymer layer interactions can also generate new coherent structures. In particular, the formation of secondary and tertiary vortices coincides with transient increases in kinetic energy, a behavior absent in the Newtonian case. Unlike classical vortex–boundary interactions, where the primary vortex survives, we find that under certain conditions it completely dissipates upon interaction with the polymer layer. These findings emphasize that fluids with non-uniform polymer concentrations, act not only as dissipative agents but also as sources of vorticity, extending the traditional view of polymer-induced drag reduction and providing new insight into vortex–polymer interactions.

\end{abstract}

\maketitle



\section{\label{sec:intro}Introduction}

\noindent Over the past few decades, the kinematics and dynamics associated with the interaction of vortical structures (e.g., vortex rings) and boundaries has received considerable attention. Such interactions underlie complex phenomena such as vortex instability, vortex reconnection, and the generation and dissipation of coherent structures. A variety of interactions have been considered such as vortex interactions with solid walls, free surfaces and contaminated surfaces.

\noindent When vortex rings impinge upon a solid boundary at normal incidence, a viscous boundary layer is produced at the boundary as a result of the no-slip condition. If the incident ring is sufficiently strong, the boundary layer separates from the wall, at which point other complex dynamics often result. Walker et al.\cite{Walker} investigated the interaction of orifice-generated vortex rings with a no-slip plane boundary to study the mechanisms responsible for the formation of secondary structures at the wall. They reported that rapid boundary layer thickening triggers a strong viscous–inviscid interaction, producing a secondary vortex ring that can interact further with the primary ring to generate a tertiary vortex. Numerical investigations such as those by Orlandi and Verzicco \cite{Orlandi_Verzicco_1993}  have identified vortex pairing as the mechanism responsible for generating secondary rings, which they found to be more unstable than the primary. Swearingen et al.\cite{Swearingen_Crouch_Handler_1995} further demonstrated that the instability of the secondary ring results from its being stretched and tilted by the strain field of the primary ring, making it highly susceptible to long-wavelength perturbations. 
\\
\noindent In contrast, interactions of vortical structures with a surface formed at the interface between two fluids such as air and water (free-surface) often differ significantly from interactions with a no-slip wall. When a vortex ring or pair impinges upon a free-surface at normal incidence, and in the absence of surface waves,  the resultant flow depends crucially on whether the surface is clean (free of contaminants), or whether surfactants are present. For a perfectly clean surface (shear-free boundary), a vortex pair does not rebound but instead follows a trajectory parallel to the interface. In contrast, with a highly contaminated surface, a viscous boundary layer forms, causing the primary vortex to rebound from the boundary. Subsequently, a secondary vortex or vortices are typically produced. This behavior has been confirmed through experiments and simulations \cite{Tsai_Yue_1995,Hirsa_Willmarth_1994, SMITH_VOLINO_HANDLER_LEIGHTON_2001}. The important observation, relevant to the present work, is that in the absence of surface waves, vortex rebound and vorticity production occurs only when the surface offers resistance to the flow.

\noindent
To summarize, the previous work associated with coherent vortical structures generated in Newtonian fluids such as vortex rings or pairs impinging on boundaries at normal incidence shows that the nature of the boundary conditions at such interfaces has a profound impact on the subsequent fluid motion. These conditions range from no-slip conditions which appear at rigid boundaries where shear stresses are high, to clean waveless free surfaces where shear is absent. For surfactant-bearing surfaces, shear stresses can vary continuously depending on the level of contamination. However, though the level of shear depends on the boundary conditions, the fluid motion in virtually all cases, with the exception of an entirely clean free surface, is qualitatively identical and can be described by the following sequence: (1) formation of a viscous boundary layer upon vortex-boundary interaction followed by (2) the rebound of the primary vortical structure, and (3) formation of other coherent vortices formed from boundary layer vorticity. We emphasize here that the primary vortex survives its interaction with the boundary. 

\noindent
In this work, we use numerical simulations to explore the interaction of a vortex pair with a concentrated region of polymeric fluid. We will emphasize the similarities and differences of the resultant fluid motion with the classical works on vortex-boundary interactions described above. In particular, we are interested in the fluid mechanics associated with the interaction of a vortex pair, which originates in a Newtonian fluid, with a polymeric fluid layer. In this case, the vortex pair encounters an abrupt change in fluid properties. As such, it cannot be easily described using standard boundary conditions (e.g., no-slip or shear-free). We emphasize that, unlike vortex-boundary interactions in Newtonian fluids in which viscous boundary fluid is the source of secondary vorticity, it will be shown that secondary vortices in vortex-polymer layer interactions are generated primarily by fluid torques generated by gradients in polymeric stresses \cite{Buckingham,estela}. We also find, in direct contrast to classical vortex-boundary interactions, that the primary vortex (in certain regions of the parameter space) is completely dissipated. Furthermore, this work involves the effects of non-uniform polymer concentrations which have recently been investigated by Han et al. \cite{Han} and Kelly et al. \cite{Kelly2024,RyanK}
  
\section{PROBLEM FORMULATION AND DIMENSIONAL ANALYSIS}

\subsection{Problem Formulation}

\noindent In this study, we investigate the motion of an incompressible, viscoelastic, non-Newtonian fluid governed by the momentum and continuity equations given by:
\begin{eqnarray}
\frac{D V_i}{Dt} &=&  {\partial_j T_{ji}} + f_i,
\label{eq:momentum}
\end{eqnarray}
and
\begin{eqnarray}
{\partial_i V_i=0},
\end{eqnarray}

\noindent where \(D{V_i}/Dt\) = \(\partial{V_i}/\partial t\) + \(V_j \partial_j V_i \), \(V_i\) = \((u,v,w)\) = \((u_1,u_2,u_3)\) are the components of the fluid velocity in the \(x,y,z\) or \(x_1,x_2,x_3\) directions, 
\(t\) is time, \( T_{ji} \) are the components of the stress tensor, and \(f_i \) are the components of the body force. Here we use a right-handed Cartesian coordinate system with the \(y\) coordinate perpendicular to the horizontal \((x-z)\) plane. The body force is used to create a vortex pair by acting impulsively on the fluid. The force is spatially distributed such that the vortex pair created by it exists solely in a horizontal \((x-z)\) plane and travels in the positive \(x\) direction. A more detailed description of the body force will be given in the numerical methods section and can also be found in \cite{Sonmez}.

\noindent The well-known FENE-P (finitely extensible nonlinear elastic-Peterlin) model \cite{polymer,tanner2000engineering} is used to represent the polymer dynamics. In this model, the components of the stress tensor \( T_{ji} \) are represented by:
\begin{eqnarray}
T_{ji} = -p{\delta_{ji}} + 2{\nu_s}S_{ji}+{\tau_{ji}}^p ,
\label{eq:stresstensor}
\end{eqnarray}

\noindent where \( p\) is the pressure divided by density, \(S_{ji}\)= \( \frac{1}{2} \) (\(\partial V_j/\partial x_i\) + \(\partial V_i/\partial x_j\)) are the components of the rate-of-strain tensor, and \(\nu_s \) is the kinematic viscosity of the solvent. The components of the polymeric stress tensor, \(\tau_{ji}^p\), are given by:
\begin{eqnarray}
{\tau_{ji}}^p=\frac{\nu_0-\nu_s}{\lambda} [f(C_{kk})C_{ji}-\delta_{ji}],
\label{eq:polymerstress}
\end{eqnarray}

\noindent where \(\nu_0 \) is the kinematic viscosity of the solution, 
\(\lambda \) is the polymer relaxation time, \(C_{ji}\) are the components of the molecular conformation tensor, and  \(f(C_{kk})\) is the Peterlin function given by:
\begin{eqnarray}
f(C_{kk})=\frac{L^2_{max}-3}{L^2_{max}-C_{kk}},
\label{eq:peterlin}
\end{eqnarray}

\noindent where \(L_{max}\) is the maximum allowable molecular extension made non-dimensional with the rest length of the polymer molecule. For later use, we define the polymer force per unit mass as \(f_{i}^p=\partial_j \tau_{ji}^p\). This expression, referring to equation \ref{eq:polymerstress}, shows that polymer forces can be especially complex as they involve gradients of the Peterlin function and the conformation tensor. In situations for which the viscosity of the fluid depends on concentration, which is the case in this work, these forces will also depend on the concentration gradients.

\noindent The polymer conformation evolves as follows:

\begin{eqnarray}
\frac{D C_{ji}}{D t} &=&  C_{jk}{\partial_k V_i} + C_{ki}{\partial_k V_j}-\frac{1}{\lambda} [f(C_{kk})C_{ji}-\delta_{ji}] + \alpha_p \nabla^2C_{ji}, 
\label{eq:conformation}
\end{eqnarray}
\noindent where the last term is the diffusion of the conformation and \(\alpha_p\) is the polymeric diffusivity, added to stabilize the numerical solution. Without this term, strong gradients in \(C_{ji}\) lead to unphysical oscillations or numerical divergence due to the convective nature of the governing equations \cite{Xi2019,dubief}. In practice, the best strategy is to select the largest polymer stress Schmidt number, \(Sc_p=\nu_s/\alpha_p\), that maintains numerical stability \cite{scp}.

\noindent The difference between the solution and solvent kinematic viscosities, \(\nu_0 - \nu_s\), depends on the concentration, \(\gamma\), which evolves according to:
\begin{eqnarray}
\frac{\partial \gamma}{\partial t}+ V_j \frac{\partial \gamma}{\partial x_j}=\alpha_m \frac{\partial^2 \gamma}{\partial x_j \partial x_j},
\label{eq:scalar}
\end{eqnarray}

\noindent where \(\alpha_m\) represents the mass diffusivity. In this work, we assume a linear relationship between the kinematic viscosity difference and the concentration given by: 
\begin{eqnarray}
\nu_0 - \nu_s = \nu_s \alpha\gamma ,
\label{eq:nuzero}
\end{eqnarray}

\noindent where \(\alpha=2.6\) x \(10^{-3} PPM^{-1}\) is an empirically determined constant reported by Nsom and Latrache \cite{Nsom} from experiments on dilute PEO (polyethylene oxide) solutions prepared in a mixture of isopropyl alcohol and water. As mentioned, we used a linear model because it provides the simplest framework capable of capturing the phenomena we are most interested in exploring. 
\noindent Using equation \ref{eq:nuzero}, the ratio of the solvent to solution kinematic viscosity,  \(\beta=  \frac{\nu_s}{\nu_0}\) , is given by \(\beta= \frac{1}{1+\alpha\gamma}\).

\subsection{Dimensional Analysis}
\noindent The objective of this work is to explore the effects of a concentrated layer of polymeric fluid (a polymer layer) on the kinematics and dynamics of a coherent region of vorticity. This is achieved by generating a vortex-pair using an impulsive body force and allowing the pair to propagate toward the layer. This problem involves not only complex vortex dynamics but also complex polymeric effects. We have, therefore, chosen to limit this complexity by solving an entirely 2-D problem in which the dynamics takes place solely in the horizontal \((x-z)\) plane. To solve this problem with the fully three dimensional code used here simply requires the body force to be independent of the \(y\) coordinate and that we also choose boundary conditions which produce a 2-D flow. For further details, see \cite{Buckingham}.
This approach requires some extra computational cost in representing the flow in the \(y\) direction, but we have found that this can be easily minimized by choosing the lowest possible resolution in that direction. 

\begin{figure}[htbp]
 \centering
 \includegraphics[width=0.95\linewidth]{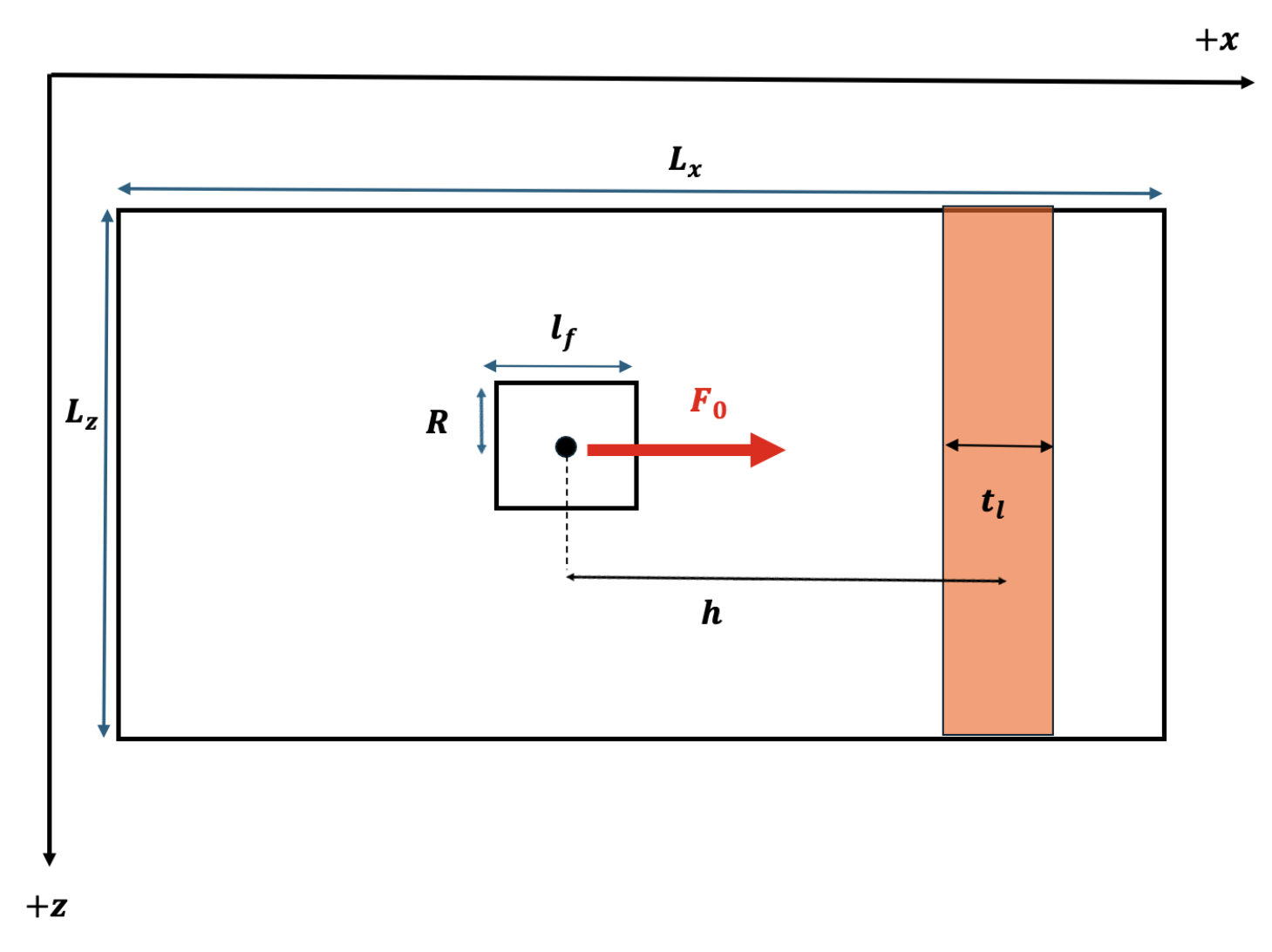}
 \caption{\label{fig:forcex} Schematic (not to scale) of the \(x-z\) plane at the center of the computational domain ( \(y=0\) ) showing the problem setup. The body force \(F_0\) acts to the right, as indicated by the red arrow,  and only within the rectangular region whose dimensions are given by \(l_f\) and \(R\). The center of this region is \(x_0,y_0,z_0\), the midpoint of the computational domain. The dimensions of the computational domain are  \(L_x \) and \(L_z\) in the \(x\) and \(z\) directions, respectively. A polymer layer of thickness \(t_l \) is positioned downstream of the body force, and \(h \) is the distance between the force center and  the initial location of the center of the polymer layer. The \(y \)-axis points outward, normal to the plane of the page. The center of the force is located at \(x=L_x/2\) and \(z=L_z/2\).}
\end{figure}

\noindent In Fig.~\ref{fig:forcex}, a horizontal \((x-z)\) region of the computational domain is shown. The dimensions of the domain are given by \(L_z\) and \(L_x\), the half-width and length of the region of fluid acted upon by the body force are given by \(R\) and \(l_f\), the distance from the center of action of the force to the initial location of the center of the polymer layer is given by \(h\), and the initial thickness of the layer is given by \(t_l\).

\noindent These geometric parameters as well as other properties of the fluid and force field can be used to determine the most important dimensionless parameters governing the problem. For example, if the volume-averaged total kinetic energy per unit mass of the system, \( E = \frac{1}{V} \int k\,dv \) where \(k=\frac{u_i u_i}{2}\) and \(V\) is the volume of the domain, is allowed to be the dependent parameter, then its dependence on the relevant physical parameters is given by:
\begin{eqnarray}
E=f(F_0, \tau, R, l_f, h, t_l, \nu_s,  \lambda,  \gamma_0,  \alpha , \alpha_p, \alpha_m, t) 
\label{eq:dimen}
\end{eqnarray}

\noindent where \(F_0\) is the body force amplitude, \(\tau\) is the temporal duration of the body force, \(\gamma_0\) is the initial maximum concentration of the polymer layer, and \(t\) is time.

If we choose a velocity scale \(u^*=\sqrt{F_0l_f}\), equation \eqref{eq:dimen} can be written in non-dimensional form as follows:
\begin{eqnarray}
\frac{E\tau^2}{l_f^2}=f(\alpha \gamma_0, W_i, \frac{t_l}{l_f}, L_{max} , Re_f, Ro , \frac{l_f}{h},\frac{R}{h}, Sc_p,Sc_m,\frac{t}{\tau})
\label{eq:nondimen}
\end{eqnarray}

\noindent Here,  \(Wi=\frac{u^* \lambda}{l_f}\) is the Weissenberg number which gives the ratio of the polymer relaxation time to a typical flow time scale, \(Re_f=\frac{u^*l_f}{\nu_s}\) is a Reynolds number characterizing the strength of the vortex pair generated by the body force, \(Ro=\frac{\tau \nu_s}{l_f^2}\) is the  Roshko number \cite{Roshko}, \(Sc_p=\frac{\nu_s}{\alpha_p} \) and  \(Sc_m=\frac{\nu_s}{\alpha_m}\) are the Schmdit numbers for the polymer stresses and polymer concentration. We have also included \(L_{max} \) in the list of non-dimensional parameters although it does not appear in dimensional form in equation \eqref{eq:dimen}. Given the large number of dimensionless numbers in this problem, we have chosen to focus on the effects of the initial polymer concentration, polymer relaxation time, polymer layer thickness, and the maximum polymer extension length, since these were found in our preliminary simulations to produce the most significant effects on the flow. Therefore, simulations were conducted in which \(\alpha\gamma_0\), \(Wi\), \(t_l/l_f\), and \(L_{max}\) were varied while keeping all other dimensionless parameters fixed. This is equivalent to varying the dimensional variables \(\gamma_0\), \(\lambda\), and \(t_l\). The fixed dimensionless numbers were: \(Re_f=\frac{u^*l_f}{\nu_s}=380.16\), \(Ro=\frac{\tau \nu_s}{l_f^2}=5.952\) x \(10^{-3}\),  \(Sc_p=\frac{\nu_s}{\alpha_p}=0.1 \), \(Sc_m=\frac{\nu_s}{\alpha_m}=4 \), \(l_f/h=0.313\) and \(R/h=0.156\). Also, \(h\), \(F_0\) and \(\nu_s\) were fixed: \(h=1.6cm\), \(F_0=100 cm s^{-1}\) and \(\nu_s=9.3\)x \(10^{-3} cm s^{-2}\).

\section{Numerical Methods}
\noindent To investigate the effects of the polymer layer on vortex dynamics, a series of numerical simulations were conducted. The governing equations  \eqref{eq:momentum}-\eqref{eq:scalar} , were solved using an in-house pseudo-spectral code \cite{transport}. The computational domain is rectangular, where the field variables are represented by Fourier modes in the horizontal plane \((x-z)\)  and  Chebyshev modes in the vertical \(y\) direction.
The domain dimensions are \(L_{x}=4 \pi\) cm in the \(x\) direction, \(L_{y}=2\) cm in the \(y\) direction, and \(L_{z}= \pi\) cm in the \(z\) direction. The grid resolution was \(512\) x  \(9\) x \(128\) in the \(x\), \(y\), and \(z\) directions.
For reasons mentioned previously, we used a fully three-dimensional code to simulate an essentially two-dimensional problem. Therefore, we used the minimum resolution in the \(y\) direction (9 grid nodes) allowed by our numerical algorithm. The corresponding grid spacing was \(\Delta x  = \Delta z =  2.454 \times 10^{-2} cm\). Grid convergence was verified by performing simulations at a higher resolution, which showed no significant change in the results. Therefore, all simulations were performed using the resolution stated above. \\
The impulsive body force used to generate a 2-D vortex pair is represented by \textbf{f} = \(F_0\)\(g(x,z)\)\(\hat{e}_x\), where  \(\hat{e}_x\) is a unit vector in the \(x\) direction and  the force distribution is given by  \(g(x,z)\). The force is maximum, \(F_0\), at the center of the domain \cite{Sonmez}. The nominal length and width of the force field are given by  \(l_f\) and \(R\) respectively where \(l_f=2R=0.5cm\) \cite{Sonmez}. To generate the vortex pair, the force acts from the start of each simulation (\(t=0\)) to a time \(t=\tau=0.16s\) at which time the force is turned off. The total time for each simulation and the time step were 1.6s and  \(\Delta t=10^{-5}s\) respectively.
In each simulation the fluid was initially at rest. The initial polymer concentration is independent of \(y\) and \(z\) and is given by the Gaussian, \(\gamma(x,t=0)=\gamma _0\ e^{-\frac{(x-x_c)^2}{2\sigma _x^2}}\), where \(x_c= (L_x/2) + h\) and the standard deviation  \(\sigma _x\) is used to define the thickness of the polymer layer, \(t_l = 2\sigma_x\).
The initial maximum polymer concentration, \(\gamma_0\), is therefore located at \(x=x_c\). All normal components of the conformation tensor were initially set to the polymer rest length (\(C_{11}=C_{22}=C_{33} = 1\)) and all other components were set to zero. 
\noindent The velocity field is subject to shear-free boundary conditions at the top and bottom walls of the domain, \( \frac{ \partial u_1}{\partial x_2}=\frac{\partial u_3}{\partial x_2}=u_2=0\), while periodic boundary conditions are applied in the \(x\) and \(z\) directions. A zero normal flux condition is enforced for all components of the conformation tensor as well as  for the polymer concentration. A total of 33 simulations were performed. Representative runs used in this study are presented in Appendix D, Table \ref{tab:table2}. All relevant symbols used throughout this work are provided in Appendix A, Table \ref{tab:table1}.

\section{Results}
\noindent We first present flow visualizations for a no polymer case and a low concentration case in order to establish a clear baseline and to highlight the differences between flows with and without polymers. The no polymer case corresponds to Run 2 in Appendix D, Table \ref{tab:table2}, while the low concentration case corresponds to Run 4 with an initial polymer concentration of \(\gamma_0=100PPM\). We then present the case which we refer to as the \textit{base case} (Run 1), chosen primarily because it illustrates a situation in which the vortex dynamics are particularly complex. 
The parameters in the base case were as follows: initial concentration, \(\gamma_0=1000 PPM\), relaxation time, \(\lambda=1s\), thickness of the polymer layer, \(t_l=2\sigma_x=\pi/5\) cm and maximum allowable molecular extension, \(L_{max}=100\). In this case, the dynamics exhibit some of the main features we wish to illustrate in this work: the ability of polymeric forces to both create and dissipate kinetic energy and vorticity. Flow visualizations are followed by results showing the effects of polymer concentration, relaxation time, polymer layer thickness, and maximum polymer extension on the evolution of the volume-averaged kinetic energy and enstrophy.

\begin{figure}[htbp!]
\centering
\includegraphics[width=0.95\textwidth, keepaspectratio, clip]{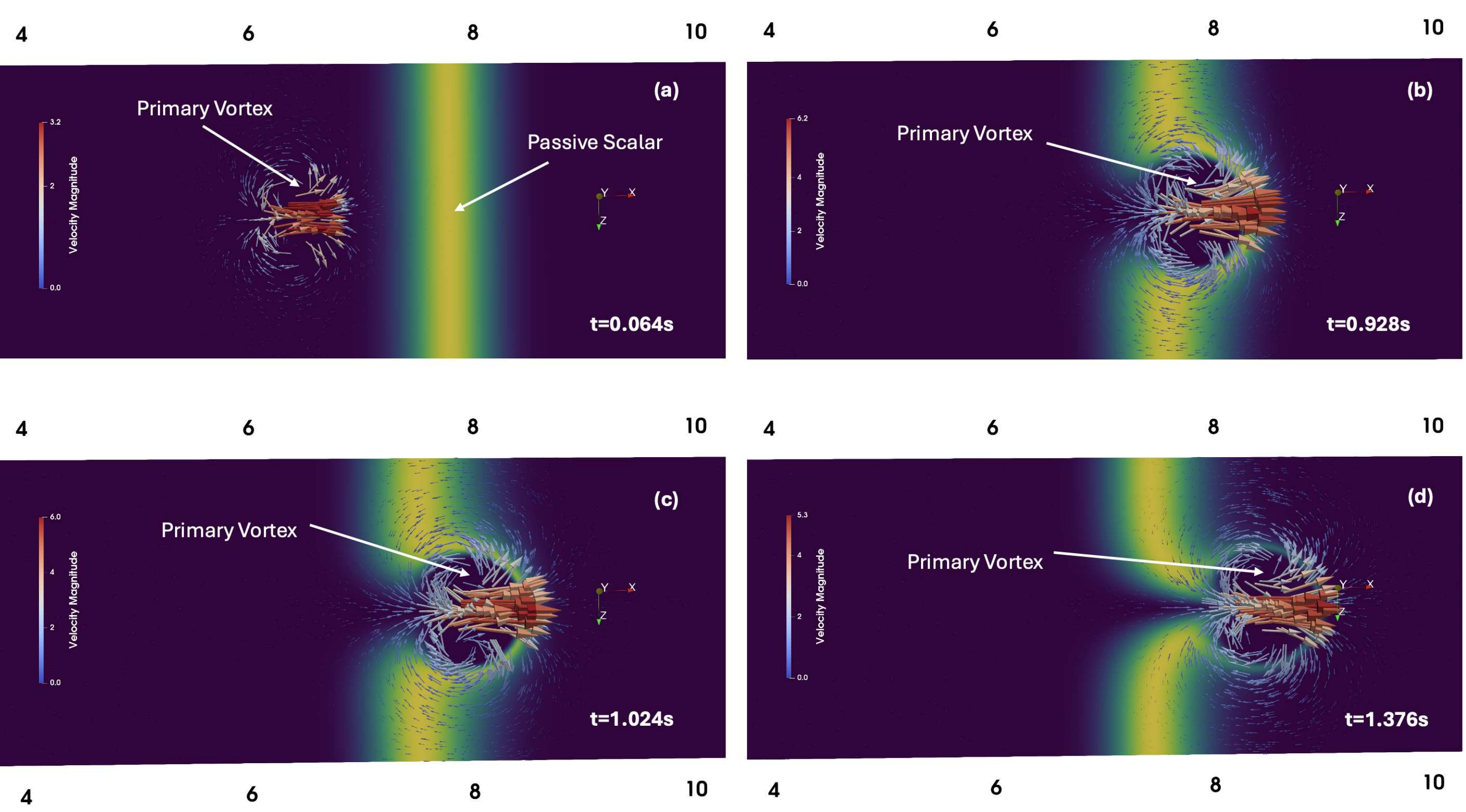}
\caption{\label{fig:passive}Evolution of a vortex pair interacting with a passive scalar (no polymer). Velocity vectors are color coded to indicate their magnitude (cm/s). The visualizations are taken from a zoomed-in region of the \(x-z\) plane at the center (\(y=0\)). The core of the primary vortex is indicated.}
\end{figure}

\begin{figure}[htbp!]
\centering
\includegraphics[width=0.95\textwidth, keepaspectratio, clip]{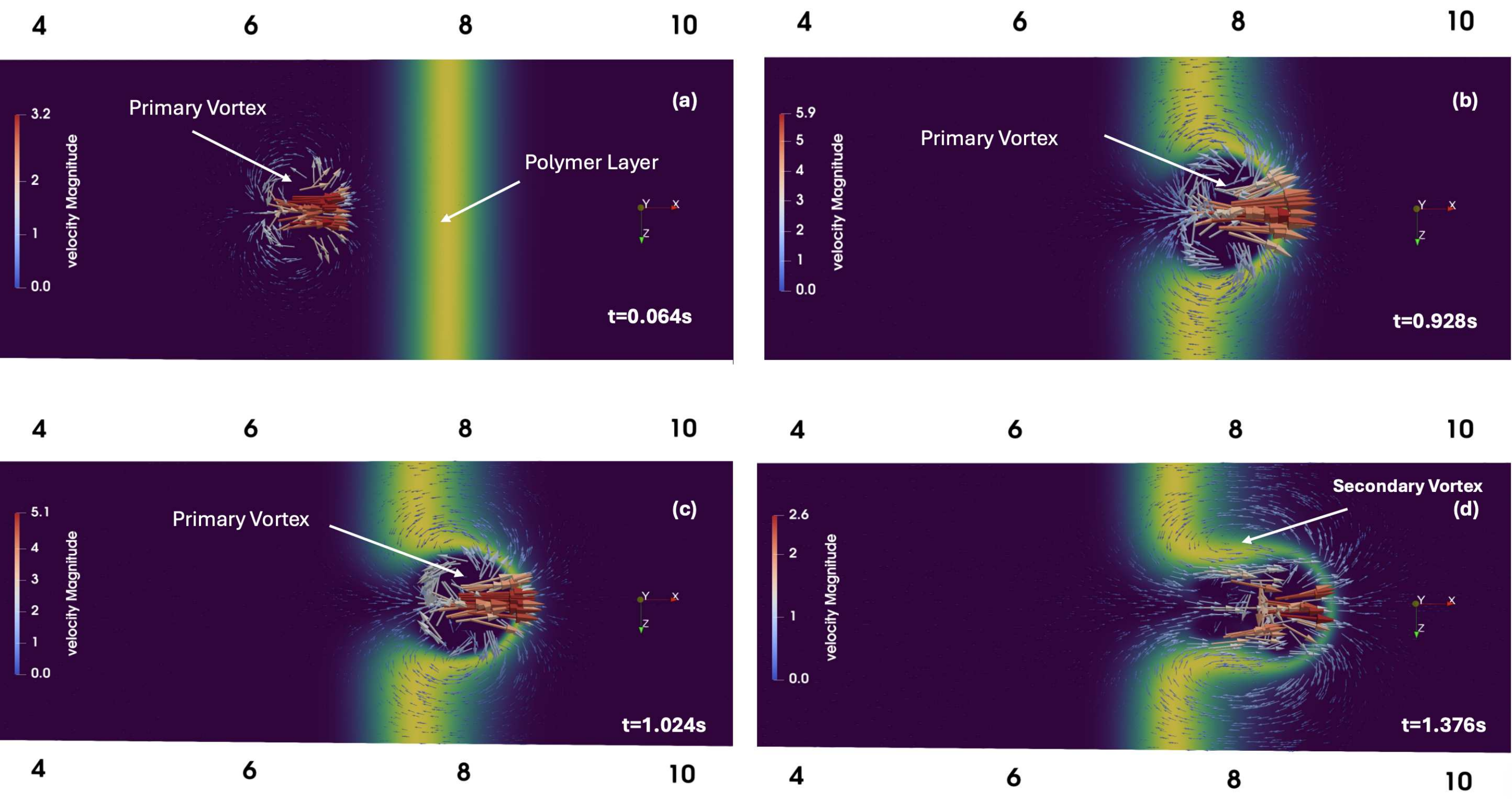}
\caption{\label{fig:con100}Evolution of a vortex pair interacting with a polymer layer for the case of \(\gamma_0=100PPM\), \(\lambda=1s\), \(t_l=\pi/5cm\), and \(L_{max}=100\). Velocity vectors are color coded to indicate their magnitude (cm/s). The visualizations are taken from a zoomed-in region of the \(x-z\) plane at the center (\(y=0\)). The core of the primary vortex as well as the formation of a secondary vortex is indicated.}
\end{figure}

\begin{figure}[htbp!]
\centering
\includegraphics[width=0.85\textwidth, keepaspectratio, clip]{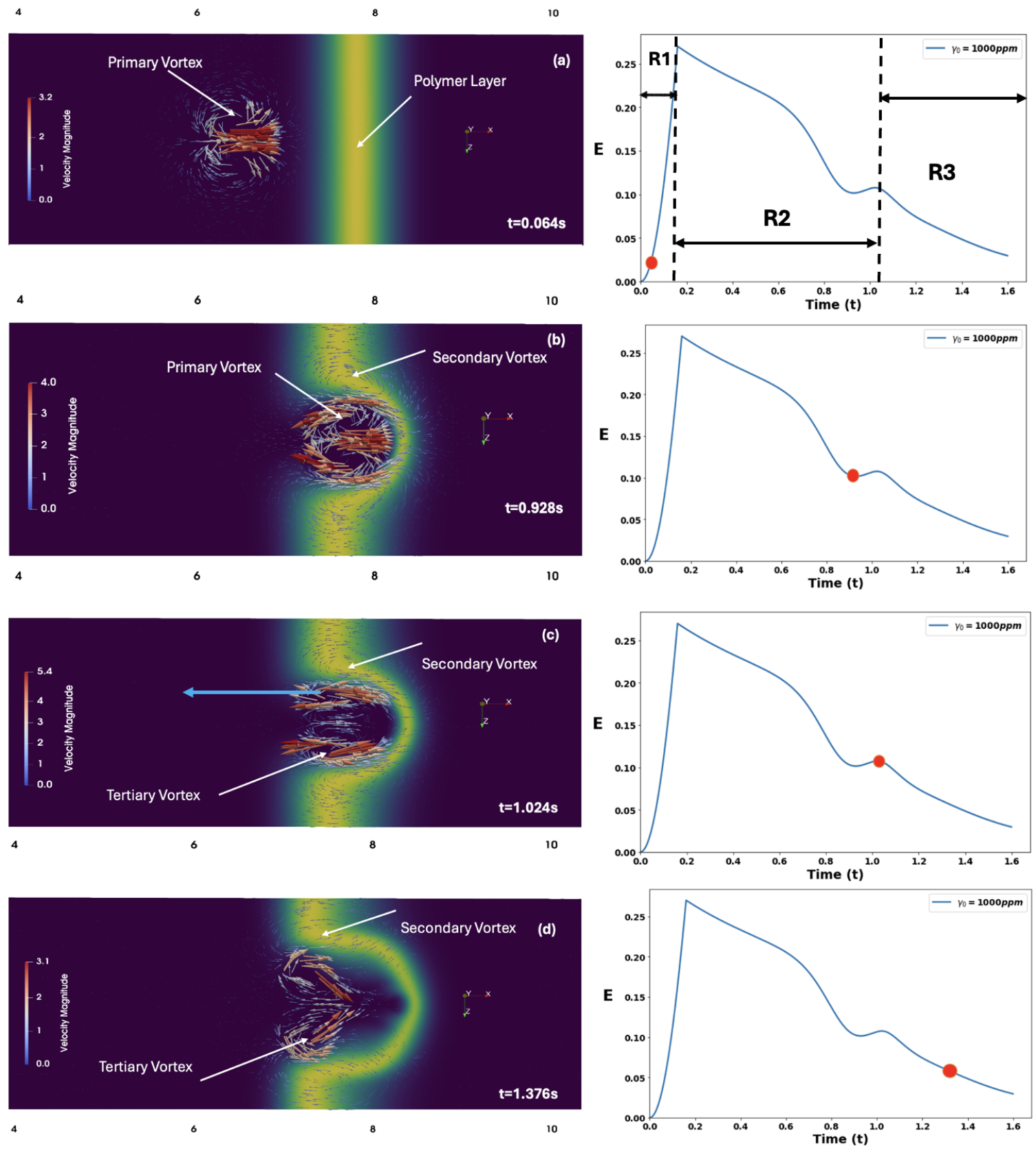}
\caption{\label{fig:vortex}Evolution of a vortex pair interacting with a polymer layer for base case where \(\gamma_0=1000PPM\), \(\lambda=1s\), \(t_l=\pi/5cm\), and \(L_{max}=100\). Velocity vectors are color coded to indicate their magnitude (cm/s). Right panels display the corresponding evolution of the kinetic energy, E. Red dots indicate the specific time instants shown corresponding to the images on the left. The visualizations are taken from a zoomed-in region of the \(x-z\) plane at the center (\(y=0\)). The core of the primary vortex as well as the formation of a secondary vortex is indicated. The blue arrow indicates the direction of translation of the tertiary vortex. In panel (a), R1 denotes the forcing phase, R2 denotes the period from force removal to the local minimum in kinetic energy, and R3 denotes the subsequent bump and decay of the kinetic energy.}
\end{figure}
\subsection{Flow visualizations}
\noindent In Fig.~\ref{fig:passive}, the evolution of a vortex pair interacting with a scalar, \(\gamma\), is shown. In this case, \(\gamma\), evolves according to equation \ref{eq:scalar} , but \(\nu_0-\nu_s=0\), and therefore the scalar is passive since all polymeric forces are turned off. However, in all other cases, unless otherwise stated, polymeric effects are turned on. At early times (Fig.~\ref{fig:passive}a), the vortex pair approaches the layer composed of the passive scalar. At later times (Fig.~\ref{fig:passive}b,Fig.~\ref{fig:passive}c and Fig.~\ref{fig:passive}d), the scalar is advected by the vortical motion but the vortex pair passes, unaffected, through the layer as seen in the video \cite{passivescalar}.

\noindent In Fig.~\ref{fig:con100}, the evolution of a vortex pair interacting with a polymer layer at a low concentration (\(\gamma_0=100PPM\)) is shown. Here, the polymer is turned off during the generation of vortex pair but it subsequently turned on. As the primary vortex interacts with the polymer layer, the primary vortex is distorted and a secondary vortex forms at later times (Fig.~\ref{fig:con100}d). This evolution is shown in the following movie \cite{conc100}.  This indicates that, at low polymer concentration, elastic effects do not significantly alter the vortex evolution leading to the examination of higher concentration cases. 

\noindent Fig.~\ref{fig:vortex} illustrates the early-time evolution of a vortex pair interacting with a polymer layer for the base case (\(\gamma_0=1000PPM\)), as defined above. In this figure, the velocity vectors as well as the location of the polymer layer are shown and a plot of the temporal evolution of the kinetic energy corresponding to each of four visualizations of the flow is presented. It is useful to divide the temporal history of the evolution of the kinetic energy into three regions: R1 (\(t<\tau=0.16s\)), R2 (\(0.16s<t<0.9s\)), and R3 (\(0.9s<t<1.6s\)). In region R1, the body force is active, but the effect of the polymer on the flow is turned off. Once the body force is turned off, polymeric effects are activated from then on. In region R2, the  kinetic energy decays monotonically until reaching a local minimum. In region R3, the kinetic energy initially increases, reaching a local maximum, before finally decaying. These regions are indicated on the  plot of kinetic energy evolution to the right of Fig.~\ref{fig:vortex}a.

\begin{figure}[htbp!]
\centering
\includegraphics[width=1.0\textwidth, trim=5 2 2 2, clip]{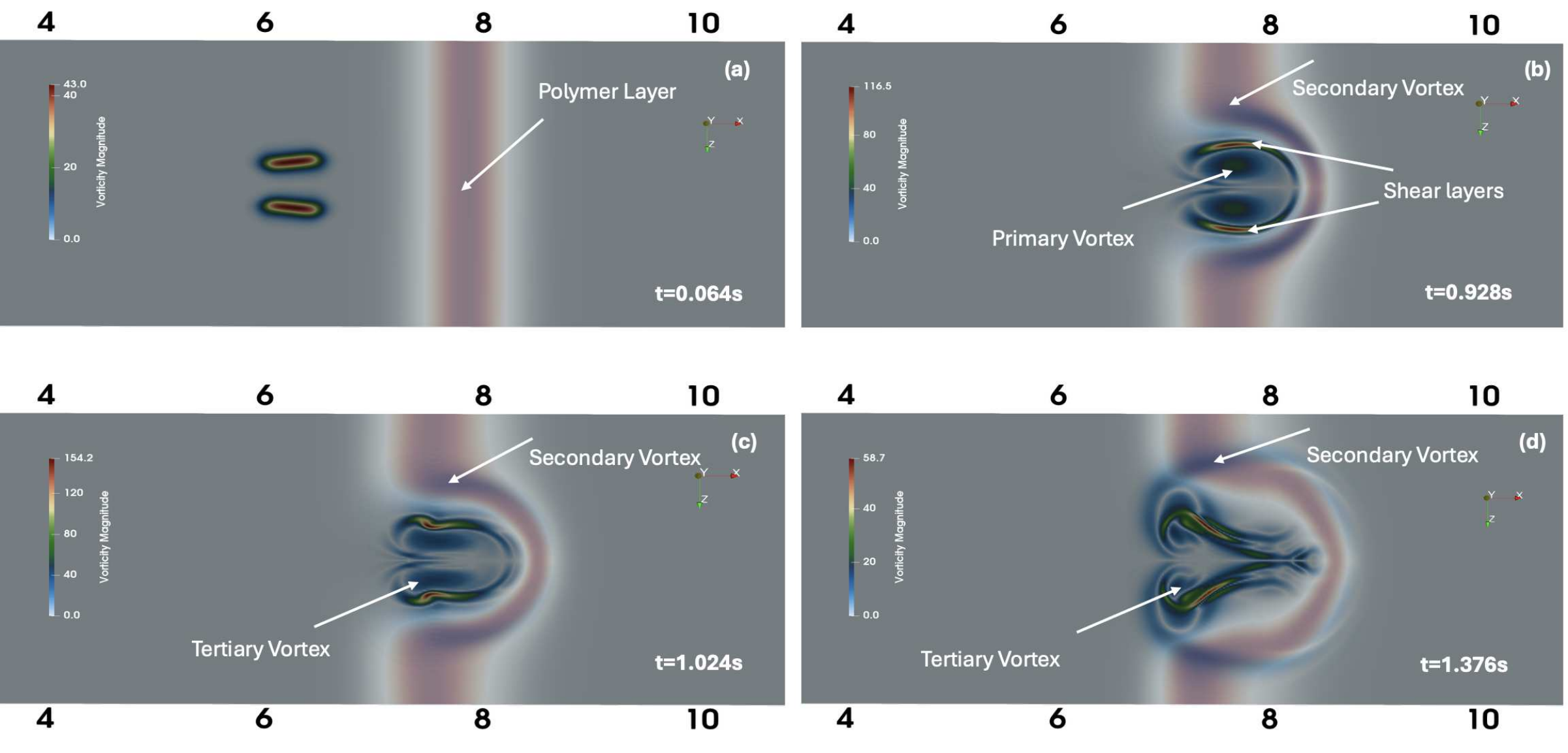}
\caption{\label{fig:shear} Evolution of the vorticity magnitude for the case shown in Fig ~\ref{fig:vortex}. Color bar represents the vorticity magnitude in \(s^{-1}\). The scalar field identifying the polymer layer is displayed with partial transparency and overlaid on the vorticity magnitude. Arrows point to the centers of the primary, secondary and tertiary vortices, as well as the shear layers in (b).}
\end{figure}

\begin{figure}[htbp!]
\centering
\includegraphics[width=1.0\textwidth, trim=5 2 2 2, clip]{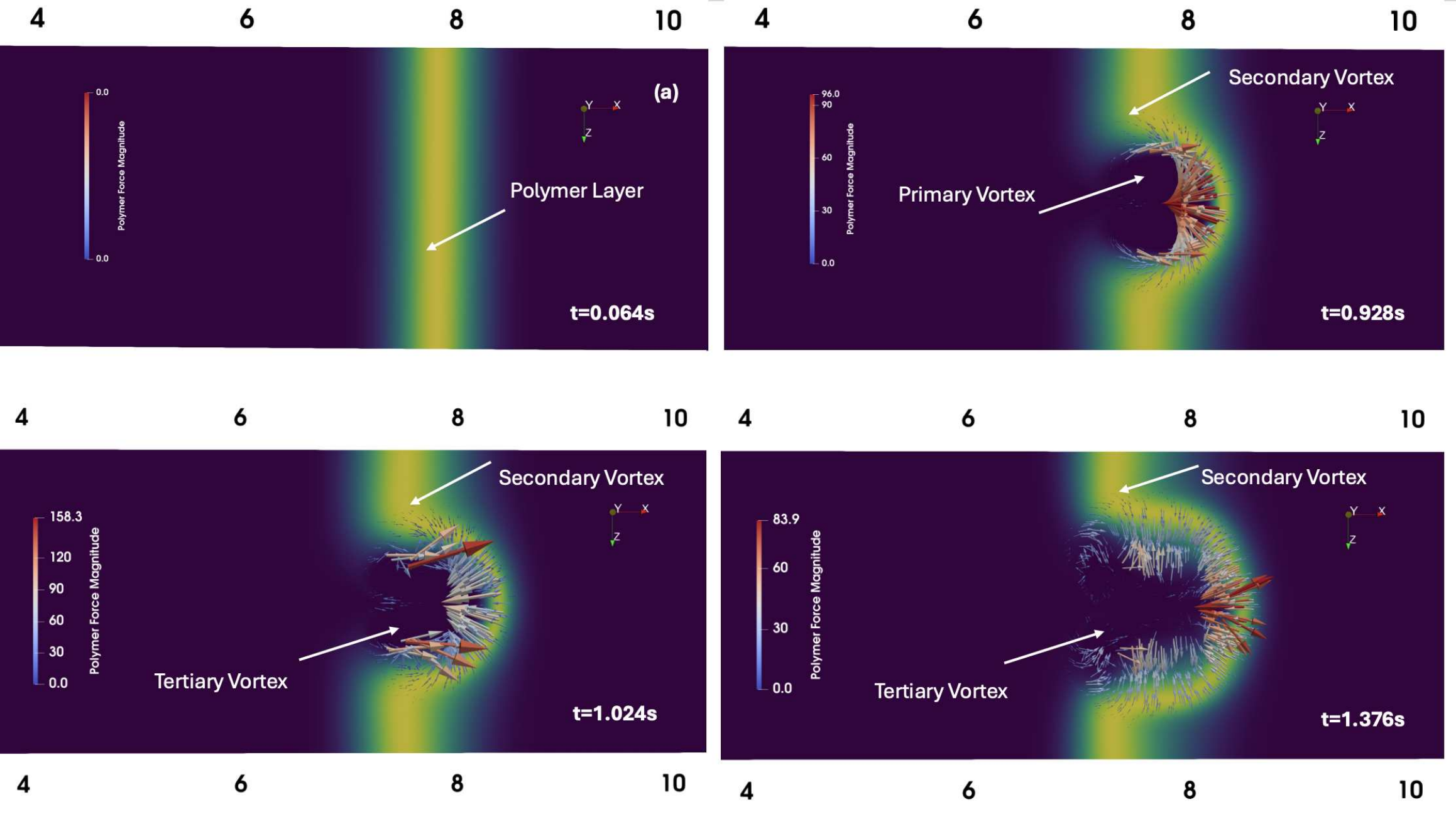}
\caption{\label{fig:force}Evolution  of the polymeric force for the case shown in Fig ~\ref{fig:vortex}. Color bars  represents the magnitude of the polymer forces per unit mass in \(cm/s^2\). Arrows indicate the magnitude and direction of the those forces.}
\end{figure}
\noindent In R1 (Fig.~\ref{fig:vortex}a), \(t=0.064s\),  the vortex pair is shown shortly after the body force has acted to create it. This primary vortex travels to right towards the polymer layer. The velocity field indicates the presence of a coherent rotational flow pattern. The kinetic energy is seen to rise sharply during the period from \(t=0 s\) to \(t=0.16 s\), in which the body force acts on the flow, and then is seen to decay after \(t=0.16 s\) when the force ceases to act. It is important to note that during this period, the fluid is completely Newtonian. Since the kinetic energy increases during this period, the rate at which work is done on the flow by the body force must be greater than the rate at which kinetic energy is being dissipated by viscous forces. This is discussed in greater detail in \cite{Sonmez}. After \(t=0.16s\) polymeric effects are turned on. 
\\
\noindent In R2, the kinetic energy is seen to decrease monotonically  (Fig.~\ref{fig:vortex}b) as the vortex pair interacts with the polymer layer. The primary vortex, which we define as the upper half of the vortex pair, is now partially embedded in the polymer layer, and a secondary vortex forms above it. The kinetic energy is seen to reach a local minimum at \(t=0.928 s\). In R3 (Fig.~\ref{fig:vortex}c, Fig.~\ref{fig:vortex}d), later stages of the vortex-layer interaction are shown. In Fig.~\ref{fig:vortex}c, at \(t=1.024s\), the primary vortex has fully entered the polymer layer, and in addition to the secondary vortex, a tertiary vortex forms. The tertiary vortex translates to the left as indicated by the blue arrow and rotates in a direction opposite to that of the primary vortex. 
\\
\noindent The kinetic energy reaches a local maximum at around \(t=1.024 s\) before continuing to monotonically decay. We will refer to this local maximum as the \textit{bump} in kinetic energy. We have observed through flow visualizations over a wide range of the parameter space that the bump in kinetic energy is always associated with the generation of the tertiary vortex.
\\
\noindent In Fig.~\ref{fig:vortex}d, at \(t=1.376s\), the primary vortex has completely dissipated, leaving only the tertiary and secondary vortices. We note that the complete dissipation of the primary vortex is in direct contrast to the case in which a vortex ring interacts with a no-slip boundary, where the primary vortex survives the interaction. The entire sequence of events for the base case shown here is given in  the following movie \cite{basecase} in which it is observed that the tertiary vortex travels to the left thereby giving the impression that it has \textit{rebounded} from the polymer layer in a manner similar to a rubber ball rebounding from a wall. 
\\
\noindent In Fig.~\ref{fig:shear}, the evolution of the vorticity magnitude in the \(x\)-\(z\) plane is presented for the base case. Arrows in Fig.~\ref{fig:shear} indicate the centers of the primary, secondary, and tertiary vortices, whose positions are consistent with those identified in the velocity field shown in Fig.~\ref{fig:vortex}. The correspondence between the locations of the peaks in the vorticity magnitude and the rotation centers in the velocity field confirms the classification of these coherent structures as distinct vortices. At early times, (Fig.~\ref{fig:shear}a), the vorticity field is characterized by two distinct vortex sheets generated by the applied body force. As the flow develops, (Fig.~\ref{fig:shear}b and Fig.~\ref{fig:shear}c) strong shear layers  emerge at the polymer-fluid interface. 
Detailed examination of movies of the flow reveals that the shear layer in the upper half of the  \(x-z\) plane is formed between the secondary and primary vortices, both of which rotate in the counter-clockwise direction. The vorticity in this layer therefore has a clockwise sense of rotation (negative vertical (y) vorticity). The clockwise rotating tertiary vortex forms rapidly due to the roll-up of this vortex layer in about 0.2 seconds after the layer forms. Referring back to Fig.~\ref{fig:vortex}b and Fig.~\ref{fig:vortex}c, this process is represented by the time taken for the kinetic energy to increase from its local minimum to its local maximum, further illustrating the correspondence between the formation of the tertiary vortex and the bump in kinetic energy. 
\\
\noindent In Fig.~\ref{fig:force}, the evolution of the polymeric vector force field (\(f_{i}^p=\partial_j \tau_{ji}^p\)) is shown. These results give further insight into the dynamics which give rise to the secondary and tertiary vortices. In Fig.~\ref{fig:force}a, these forces are absent since the polymer concentration is nearly zero in regions where flow is present. Later in time, (Fig.~\ref{fig:force}b, Fig.~\ref{fig:force}c and Fig.~\ref{fig:force}d), polymeric forces appear primarily at the edge of the polymer layer. These forces clearly induce a  clockwise torque on the fluid in the upper region of the flow and a counter-clockwise torque in the lower region. These torques are in fact exactly what is required to generate the clockwise rotating tertiary vortex in the upper region and counter-clockwise tertiary vortex in the lower region shown in Fig.~\ref{fig:vortex}. We can therefore refer to these torques as counter-torques, since they can generate vorticity whose sign is opposite to that of the primary vortex. Finally, as shown in Fig.~\ref{fig:force}d, the force field progressively weakens as the flow decays.
\\
\noindent In summary, these flow visualizations show that, unlike flows associated with vortices interacting with no-slip walls and free surfaces, vortex-polymer layer interactions generate new vortices coincident with the disappearance of the primary vortex. In additon, we show that polymer torques coincide with the generation of new vortical structures.

\begin{figure}[htbp!]
\centering
\includegraphics[width=0.6\textwidth, trim=2 2 2 2, clip]{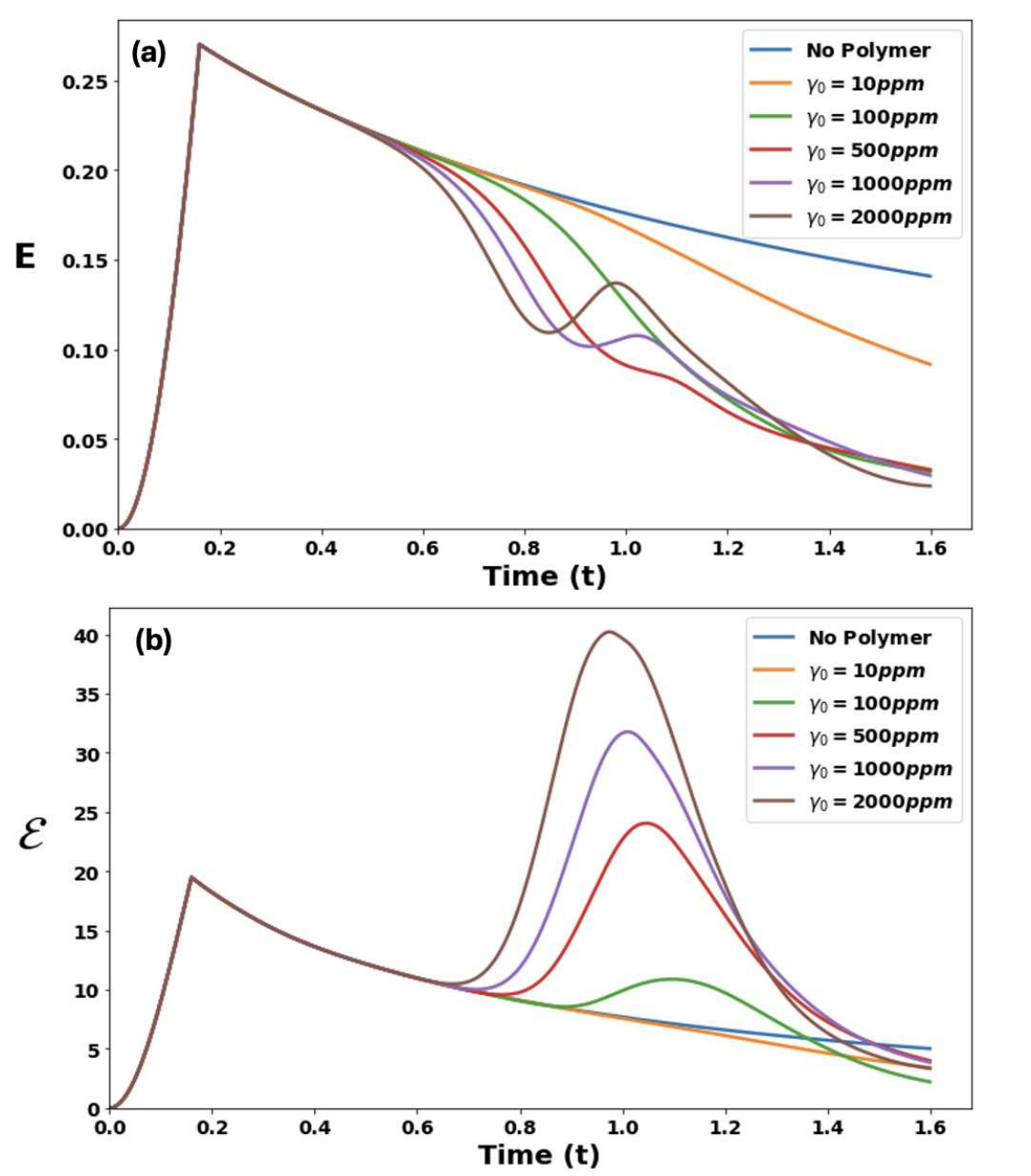}
\caption{\label{fig:conce}a) Evolution of the total kinetic energy per unit mass, E, ( \(cm ^{2}s^{-2}\) ) vs time in seconds for a range of polymer concentrations.  b) Evolution of the enstrophy,  \(\varepsilon\) (\(s^{-2}\)), for various polymer concentrations. The polymer layer thickness, relaxation time, and maximum polymer extension are held fixed at \(t_l=\pi/5 cm\), \(\lambda=1s\) and \(L_{max}=100\) respectively, while the polymer concentration is varied.}
\end{figure}

\subsection{Evolution of kinetic energy and enstrophy}
\begin{figure}[hbt!]
\centering
\includegraphics[width=0.6\textwidth, trim=0 2 2 2, clip]{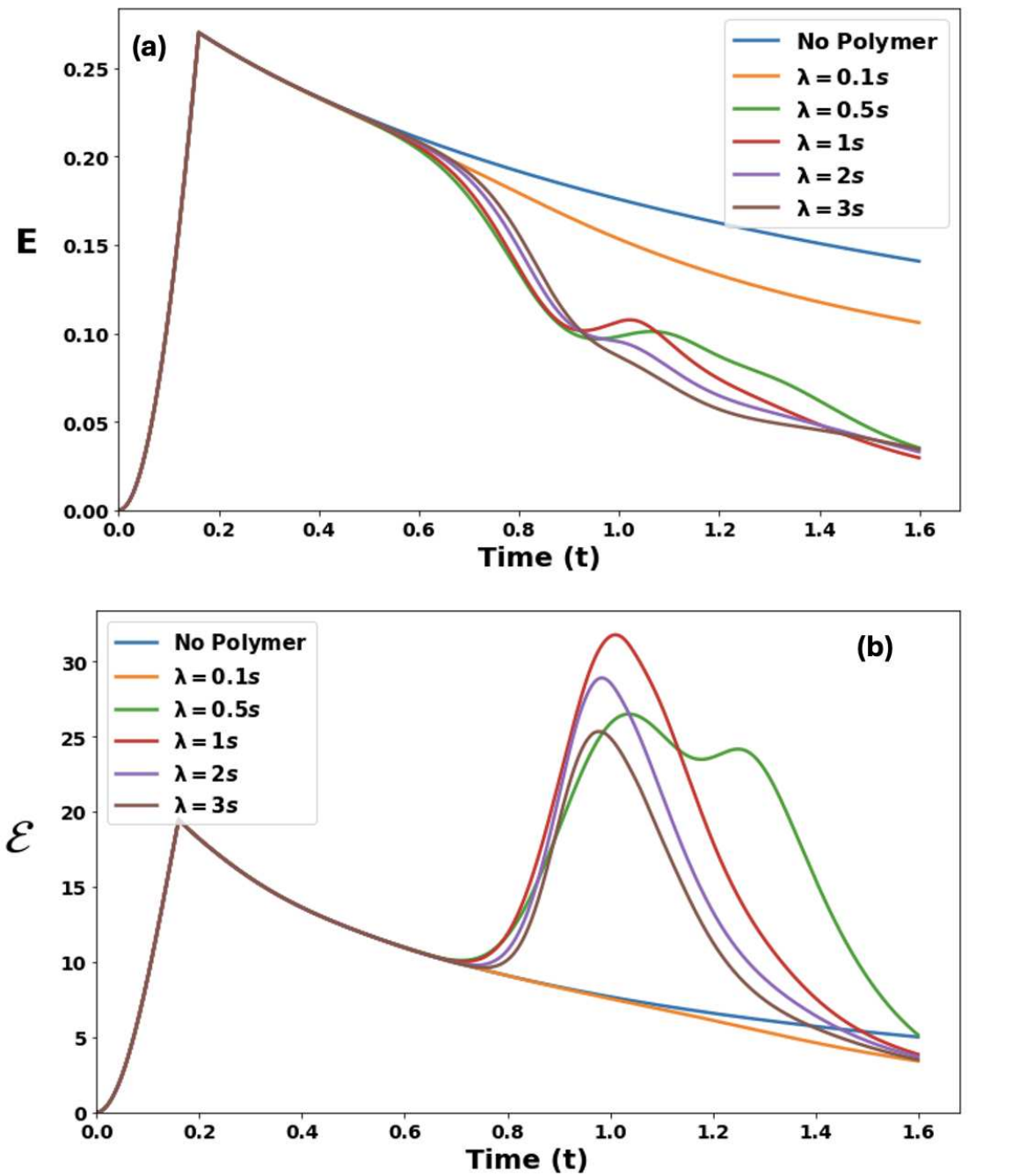}
\caption{\label{fig:lambda}a) Evolution of total kinetic energy per unit mass, E, for varying polymer relaxation times, \(\lambda\). b) Corresponding enstrophy evolution. Polymer layer thickness, polymer concentration, and maximum polymer extension are held fixed at \(t_l=\pi/5 cm\), \(\gamma_0=1000PPM\) and \(L_{max}=100\) while the polymer relaxation time is varied.}
\end{figure}

\begin{figure}[htbp!]
\centering
\includegraphics[width=0.6\textwidth, trim=2 2 2 2, clip]{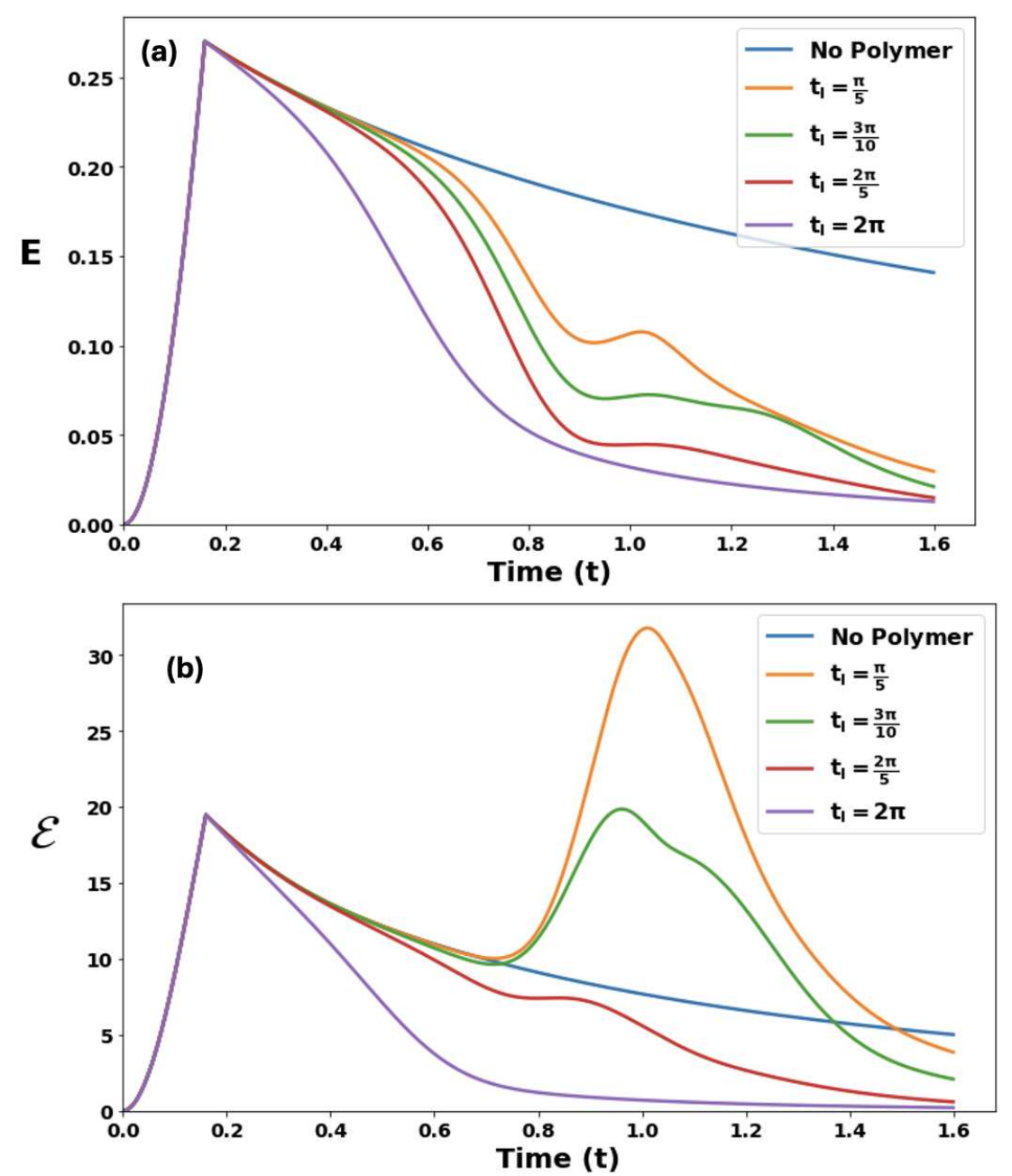}
\caption{\label{fig:tl}a) Evolution of the total kinetic energy per unit mass for a range of  polymer layer thicknesses, \(t_l\). b) Corresponding evolution of the enstrophy, \(\varepsilon\). Relaxation time, polymer concentration, and polymer length are held fixed at \(\lambda=1s\), \(\gamma_0=1000PPM\) and \(L_{max}=100\), while the layer thickness \(t_l\) is varied.}
\end{figure}

\begin{figure}[htbp!]
\centering
\includegraphics[width=0.6\textwidth, trim=2 2 2 2, clip]{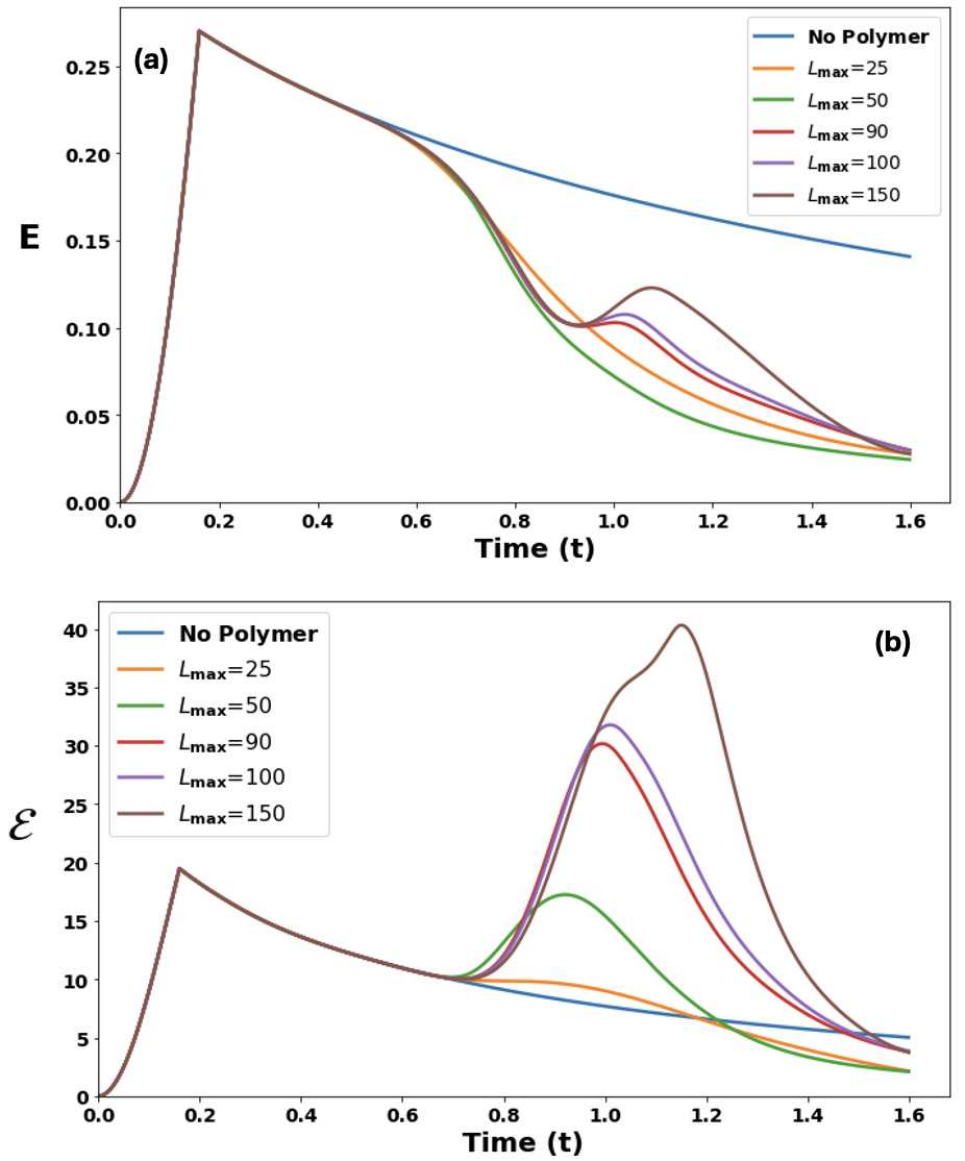}\caption{\label{fig:lmax}a) Evolution of total kinetic energy per unit mass, E, for varying \(L_{max}\) vs time in seconds  b) Corresponding evolution of enstrophy, \(\varepsilon\). Relaxation time, polymer concentration, and layer thickness are fixed at \(\lambda=1s\), \(\gamma_0=1000PPM\) and \(t_l=\pi/5cm\) while the polymer extension is varied.}
\end{figure}

\noindent In Figs.~\ref{fig:conce}-~\ref{fig:lmax} the effects of polymer concentration, relaxation time, polymer layer thickness, and maximum polymer extension on the evolution of the volume-averaged kinetic energy and enstrophy are shown. In each of these figures, one of these four parameters is varied while the other three are held fixed. Here, we define the volume-averaged enstrophy as \(\varepsilon=\frac{1}{V} \int \epsilon dV\) where \(\epsilon=w_iw_i\), and \(w_i\) represents the components of the vorticity. For reference purposes, we provide derivations of the volume-averaged enstrophy and the kinetic energy per unit mass in Appendices B and C, respectively. 

\noindent In Fig.~\ref{fig:conce}, layer thickness, relaxation time, and maximum polymer extension are fixed (  \(\lambda=1s\), \(t_l=\pi/5 cm\), and \(L_{max}=100\) ) while polymer concentration is varied. In Fig.~\ref{fig:conce}a, in the case of no polymer (\(\gamma=0\)), we see that after the force ceases to act, the total kinetic energy drops monotonically. This is also the case for the lowest concentrations (\(\gamma_0=10 PPM\)) and (\(\gamma_0=100 PPM\)). However, for \(\gamma_0=500 PPM\) a bump in kinetic energy first appears, with its amplitude reaching a peak at \(\gamma_0=2000 PPM\), the highest concentration examined in this work. It is important to note that in no case in which polymer is present does the kinetic energy ever exceed that of the Newtonian case. Futhermore, the rate of decay of the kinetic energy in the time interval \(0.6s < t < 0.9s\) is generally greater, and in some cases significantly greater, when polymer is present compared to the no polymer case. 

\begin{figure}[htbp!]
\centering
\includegraphics[width=0.6\textwidth, trim=2 2 2 2, clip]{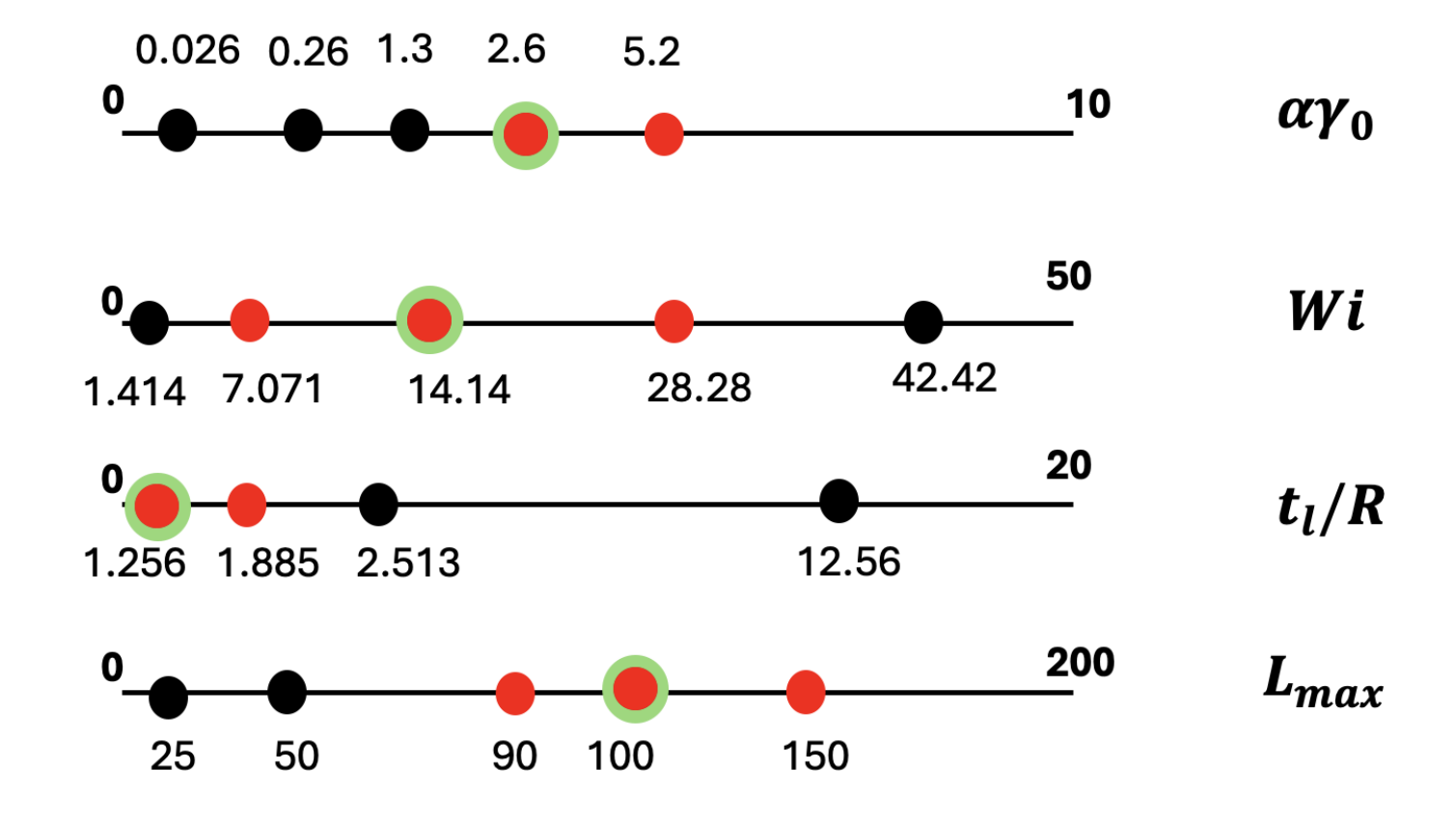}\caption{\label{fig:nondim}Non-dimensional parameter ranges. Each horizontal line represents the range of non-dimensional parameters examined. Red markers indicate cases in which a bump in kinetic energy was observed, while black markers correspond to cases without such behavior. Green-circled red markers denote the base case, which also exhibits a bump in kinetic energy.}
\end{figure}

\noindent The enstrophy evolution, however, exhibits a different behavior (Fig.~\ref{fig:conce}b). Here, we observe that in all cases in which polymer is present, with the exception of the lowest polymer concentration, that the global enstrophy increases substantially above the Newtonian case. As the vortex pair interacts with the polymer layer, strong velocity gradients are generated as the elastic layer resists fluid motion. These strong gradients correspond directly to the large increases in enstrophy shown in Fig.~\ref{fig:conce}b. The enstrophy maxima increase monotonically as concentration increases. We will refer to a local maximum in enstrophy as an enstrophy bump.
\\
In Fig. \ref{fig:lambda}, polymer concentration, layer thickness and maximum polymer extension are fixed ( \(\gamma_0=1000PPM\), \(t_l=\pi/5 cm\),  and \(L_{max}=100\) ) while the polymer relaxation time is varied.
\noindent The kinetic energy in both the no polymer case and the lowest relaxation time (\(\lambda=0.1s\)) case exhibits a monotonic decline once the body force is removed (Fig.~\ref{fig:lambda}a). We observe that the bump in kinetic energy generally decreases with increasing \(\lambda\), except for the case \(\lambda=0.5s\). Interestingly, the \(\lambda=0.5s\) case exhibits two small bumps in kinetic energy, which are also reflected in the evolution of the enstrophy in Fig.~\ref{fig:lambda}b. At present, the origin of this behavior remains unclear. The enstrophy is seen to reach its maximal value in the base case, and this occurs simultaneously with the bump in kinetic energy. The peak value of the enstrophy is seen to decrease as the relaxation time increases from \(\lambda=1s\) to \(\lambda=3s\)
\\
\noindent In Fig.~\ref{fig:tl}, the relaxation time, polymer concentration, and polymer length are fixed ( \(\gamma_0=1000PPM\), \(\lambda=1s\), and \(L_{max}=100\) ) while the layer thickness \(t_l\) is varied. In Fig.~\ref{fig:tl}a 
(R2), we see that the rate of decay in kinetic energy increases as the layer thickens. In R3 a noticeable bump in kinetic energy is seen in the base case \(t_l=\pi/5\) at \(t=1.1s\).
However, the influence of \(t_l\) on enstrophy is more complex compared to Fig.~\ref{fig:conce} and Fig.~\ref{fig:lambda}. Here we find that at smaller thicknesses (\(t_l=\pi/5\) and \(t_l=3\pi/10\)), the enstrophy increases significantly above the Newtonian case, while at larger thicknesses (\(t_l=2\pi/5\) and \(t_l=2\pi\)) it falls below it.
\\ 
In Fig.~\ref{fig:lmax}, the polymer concentration, relaxation time, and layer thickness are fixed (\(\gamma_0=1000PPM\), \(\lambda=1s\), and \(t_l=\pi/5cm\) ) while the polymer maximal extension length is varied.
\noindent In Fig.~\ref{fig:lmax}a, the kinetic energy bump is highest for the largest polymer length and decreases with \(L_{max}\). No bump in kinetic energy is observed for \(L_{max}=25\) and \(L_{max}=50\). In
Fig.~\ref{fig:lmax}b, we observe that as \(L_{max}\) increases, the enstrophy rises above the Newtonian case.
\\
\noindent In Table \ref{tab:table2}, Appendix D, we give additional details regarding the conditions under which a tertiary vortex forms. In the last three columns of Table \ref{tab:table2}, the presence (Yes) or absence (No) of a kinetic energy bump, an enstrophy bump, and the formation of a tertiary vortex is shown. These results show that if a bump in kinetic energy and enstrophy are observed in a given case, then a tertiary vortex is always present. However, there are some cases in which a tertiary vortex can form when a bump appears in either kinetic energy or enstrophy. Furthermore, we find that in all cases where a tertiary vortex is present, a secondary vortex is always observed.
\\
\noindent These parametric studies highlight the sensitive dependence of vortex-polymer interactions on polymer concentration, relaxation time, polymer layer thickness, and maximum polymer extension. While the kinetic energy never exceeds the Newtonian case, the appearance and amplitude of transient bumps indicate that polymers modify the dynamics of momentum transfer and dissipation. Importantly, the appearance of bumps in kinetic energy and enstropy is always associated with the generation of tertiary vortices as well as the dissipation of the primary vortices.
However, when bumps are not observed, the primary vortex survives the interaction. It is also observed that enstrophy is consistently amplified in the presence of polymers, reflecting the strong velocity gradients generated as the elastic layer resists vortex-induced deformation. More specifically, we expect that the interaction between a vortex pair and a polymer layer to be strongest when polymer forces are large. Examination of the expression for polymer forces,
\(f_{i}^p=\partial_j \tau_{ji}^p\), together with equations \eqref{eq:polymerstress} and \eqref{eq:peterlin}, shows that polymer forces increase when polymer stresses and their spatial gradients increase. It follows that increasing polymer concentrations, and  \(L_{max}\) will lead to larger \(f_{i}^p\). Furthermore, for a fixed concentration, decreasing polymer layer thickness leads to larger stress gradients and thus to larger polymeric forces. These arguments are supported by the results shown in Fig.~\ref{fig:nondim}, which summarizes the effects on the flow associated with the non-dimensional parameters considered in this study. Red markers indicate cases in which a distinct bump in kinetic energy is observed, while black markers correspond to cases with no such bump in kinetic energy. Green-circled red markers denote the base case. These results show that the strongest interactions occur for polymer concentrations and \(L_{max}\) values above certain thresholds, whereas only interactions with thin polymer layers exhibit bumps in kinetic energy. It is also evident that a bump in kinetic energy is observed only within a range of Weissenberg numbers.

\section{Discussion}
\noindent In this work, we investigated the interaction of a vortex pair with a localized layer of polymeric fluid of non-uniform concentration. Previous research on vortex interactions with boundaries and interfaces in Newtonian fluids demonstrated that boundary conditions, such as no-slip, shear free, or conditions associated with surfactants, fundamentally alter the flow dynamics. Such studies have consistently shown that the primary vortex typically survives the interaction, while secondary vortices are generated from boundary layer fluid.
\\
Our study extends this classical understanding by showing that the presence of polymers modifies the interaction in two different ways. First, polymeric stresses not only dissipate vorticity, as reported in the drag-reduction literature, but also act as a source of new coherent structures \cite{Lumley1969,Virk1975}. In particular, the current investigation has shown that gradients in polymeric stresses produce fluid torques that are directly responsible for the generation of secondary and tertiary vortices. This demonstrates, consistent with the work of de Genes and Min et al.\cite{DEGENNES19869, MIN_YUL}, that polymers can both extract energy from the flow and inject new vorticity into it. Second, while most prior drag-reduction studies have focused on so-called \textit{polymer oceans}, in which the polymer is uniformly distributed throughout the fluid \cite{Virk1975,Lumley1969},
our work emphasizes the effects of a non-uniform polymer concentration layer. This distinction is crucial because the localized nature of the polymer layer creates sharp gradients in elastic stresses that amplify vorticity and trigger the generation of new coherent vortices. The complete dissipation of the primary vortex in certain parameter regimes further underscores the departure from classical Newtonian vortex–boundary interactions, in which the primary structure typically survives.

\noindent Finally, it is interesting to note a possible connection between the results of this work and the phenomenon of drag reduction. Serafini et al.\cite{SERAFINI2023104471} simulated polymeric turbulent pipe flow using a Lagrangian FENE-P model, and defined a  Weissenberg number using the outer variables \(U_b\) (bulk velocity in the pipe) and \(r_p\) (pipe radius) as  \(Wi_s=U_b \lambda/r_p\). They observed that drag reduction appeared first for \(Wi_s \approx 1\) and  reached approximately 25\% at \(Wi_s \approx 100\). This aligns with our results which show maximum vortex-layer interaction in the same Weissenberg number range, where we note that our Weissenberg number is also defined using the outer variables like \(u^*\) and \(l_f\). This alignment, though intriguing, should not be overinterpreted. On the other hand, the drag reduction phenomenon occurs when the polymer relaxation time is on the order of the characteristic flow time scale \cite{Lumley1973}. This interpretation is consistent with the fact that we observe strong interactions only within a specific range of \(Wi\) reinforcing the idea that time scale matching between the flow time scale and the polymer relaxation time is central to the dynamics observed in this work.
\begin{acknowledgments}
\noindent Support was provided by the National Science Foundation under Grants No. 1904953 and No. 1905288. Simulations were performed on ARGO, a research computing cluster provided by
the Office of Research Computing at George Mason University. Additional support for R.K. was provided by the National Science Foundation Graduate Research Fellowship under Grant No. DGE 2137420.
\end{acknowledgments}

\appendix
\section{TABLE OF SYMBOLS}
Table \ref{tab:table1} contains all of the relevant symbols used in this work.
\begin{table}[H]
\caption{\label{tab:table1}%
This table summarizes all symbols used in this work. For each entry, a definition is provided along with an indication of whether the quantity is dimensional (e.g., centimeters, seconds) or nondimensional.
}
{\scriptsize
\begin{ruledtabular}
\begin{tabular}{ccc}

\textrm{Symbol}&
\textrm{Definition}&
\textrm{Units}\\
\colrule
$x,y,z$ & Coordinates & Dimensional \\
$x_1, x_2,x_3$ &Scripted coordinates corresponding to $x, y,z$ & Dimensional \\
$x_0, y_0,z_0$ &Midpoint of the computational domain & Dimensional \\
$L_x,L_y,L_z$ & Domain dimensions in $x,y,z$ & Dimensional \\
$t$ & Time & Dimensional \\
$V_i$ & Components of the velocity & Dimensional \\
$u,v,w$ & Components of the velocity in the x, y, and z directions & Dimensional \\
$u_1,u_2,u_3$ & Scripted components of the velocity & Dimensional \\
$p$ & Pressure & Dimensional \\
$ \rho $ & Density & Dimensional \\
$T_{ij}$ & Components of stress tensor & Dimensional\\
$F_0$ & Body force amplitude & Dimensional \\
$S_{ij}$ & Components of rate of strain tensor & Dimensional \\
$ \tau_{ij}^p $ & Components of polymeric stress tensor & Dimensional \\
$ f_i^p$ & Polymer force per unit mass & Dimensional \\
$\nu_0$ & Kinematic viscosity of the solution & Dimensional \\
$\nu_s$ & Kinematic viscosity of the solvent & Dimensional \\
$\lambda$ & Relaxation time of the polymer & Dimensional \\
$C_{ij}$ & Components of the molecular conformation tensor & Dimensional \\
$L_{max}$ & Maximum allowable molecular extension & Nondimensional \\
$\alpha_p$ & Polymer diffusivity & Dimensional \\
$\alpha_m$ & Mass diffusivity & Dimensional \\
$\alpha$ & Empirically determined constant & Dimensional \\
$R$ & Body force radius & Dimensional \\
$h$ & Distance from center of the body force to center of the polymer layer & Dimensional \\
$l_f$ & Body force length & Dimensional \\
$\gamma$ & Polymer concentration & Dimensional \\
$\gamma_0$ & Initial maximum polymer concentration & Dimensional \\
$u^*$ & Characteristic velocity associated with the vortex pair & Nondimensional \\
$\tau$ & Body force duration & Dimensional \\
$Wi$ & Wiessenberg Number & Nondimensional \\
$Re_f$ & Force based Reynolds Number & Nondimensional \\
$Ro$ & Roshko number & Nondimensional \\
$Sc_p$ & Schmidt number for polymer & Nondimensional \\
$Sc_m$ & Schmidt number for polymer concentration & Nondimensional \\
$E$ & Volume- averaged kinetic energy per unit mass & Dimensional \\
$\Delta x,\Delta z$ & Computational grid resolution in x and z directions & Dimensional \\
$\omega_i$ & Components of the vorticity & Dimensional\\
$\varepsilon$ & Volume-averaged enstrophy & Dimensional\\
\end{tabular}
\end{ruledtabular}
}
\end{table}

\section{Derivation of Enstropy Equation }
\noindent Here, we outline the derivation of 
the equation for the evolution of the volume-averaged enstrophy,  \(\varepsilon\), for a viscoelastic fluid. The volume-averaged enstrophy  is obtained by taking the curl of the momentum equation \eqref{eq:momentum} which yields the vorticity transport equation. The inner product of this equation with the vorticity vector is then taken, followed by integration over the entire computational domain. By applying the divergence theorem and enforcing the boundary conditions, surface integrals vanish, leaving only volumetric contributions. The resulting expression for the rate of change of volume-averaged enstrophy is

\begin{equation}
    \frac{d \varepsilon}{dt} 
    = - \nu_s \overline{\left(\partial_j \omega_i\right)^2} 
    + \overline{\omega_i Q_i},
\end{equation}

\noindent where \(\nu_s\) denotes the solvent kinematic viscosity, \(\omega_i\) is the vorticity, and \(Q_i\) is the torque associated with the polymeric force. The overbar \(\overline{(.)}\) represents a volumetric average over the entire computational domain. The first term on the right-hand side corresponds to the viscous dissipation of enstrophy, while the second term represents the enstrophy production by polymeric torques.

\section{Derivation of Kinetic Energy Equation}
\noindent Here, we outline the derivation of 
the equation for the evolution of the volume-averaged kinetic energy per unit mass, \(E\), for a viscoelastic fluid. Multiplying \eqref{eq:momentum} by the velocity component \(V_i\), integrating over the computational domain, and applying the boundary conditions yields;

\begin{equation}
    \frac{d E}{dt} 
    = - \nu_s \overline{\left(\partial_j V_i\right)^2} 
    + \overline{V_if_i^p}+\overline{V_if_i}.
\end{equation}

\noindent The first term on the right-hand side corresponds to viscous dissipation, the second term represents the work done by polymeric forces on the flow, and the third term represents the work done by external body forces. 

\section{SIMULATION PARAMETERS}
\noindent Table \ref{tab:table2} lists the parameters used in all the simulations. Initial maximum polymer concentration, \(\gamma_0\), polymer relaxation time, \(\lambda\), thickness of the polymer layer, \(t_l\), non-dimensional concentration, \(\alpha\gamma_0\), Wiessenberg number, \(Wi\), ratio of the layer thickness to force nominal length, \(t_l/l_f\), and the maximum polymer length, \(L_{max}\) are listed. 

\begin{table}[h]
\setlength{\tabcolsep}{4pt}
\caption{\label{tab:table2}%
This table lists the parameters used in all simulations.
}
{\small
\begin{ruledtabular}
\begin{tabular}{l|ccc|cccc|ccc}
\textrm{Run \#} &
\textrm{$\gamma_0$ (ppm)} &
\textrm{$\lambda$ (s)} &
\textrm{$t_l$ (cm)} &
\textrm{$\alpha \gamma_0$}&
\textrm{$Wi$} &
\textrm{$t_l/l_f$} & 
\textrm{$L_{max}$} &
\makecell{\textrm{Bump in}\\$E$} &
\makecell{\textrm{Bump in}\\$\varepsilon$} &
\makecell{\textrm{Tertiary}\\Vortex}\\
\colrule
1 (Base case)  & 1000   &1.0 & $\pi$/5   & 2.6  & 14.14  &1.256 & 100 & Yes & Yes & Yes \\

2 (No polymer)  & -  & -   & -  & -  & -  & - & -& No & No & No \\

3 & 10   & 1.0 & $\pi$/5 & $2.6 \times 10^{-2}$ & 14.14   & 1.256 & 100 & No & No & No  \\

4 & 100  & 1.0 & $\pi$/5  & $2.6 \times 10^{-1}$ & 14.14  & 1.256 & 100 & No & Yes & No \\ 

5 & 500  & 1.0 & $\pi$/5  & 1.3  & 14.14   & 1.256 & 100 & No & Yes & Yes \\

6 & 2000 &1.0  &$\pi$/5   & 5.2  & 14.14   & 1.256 & 100 & Yes & Yes & Yes \\

7 & 1000 & 0.1 & $\pi$/5  & 2.6  & 1.414   & 1.256 & 100 & No & No & No \\

8 & 1000 & 0.5 & $\pi$/5  & 2.6  & 7.071  & 1.256 & 100& Yes & Yes & Yes \\

9 & 1000 & 2.0 & $\pi$/5  & 2.6  & 28.28  & 1.256 & 100& No & Yes & Yes  \\

10& 1000 & 3.0 & $\pi$/5  & 2.6  & 42.42    & 1.256 & 100 & No & Yes & Yes  \\

11 & 1000 & 1.0 & $3\pi$/10 & 2.6  & 14.14   & 1.885 & 100 & No & Yes & Yes \\

12 & 1000 & 1.0 & $2\pi$/5  & 2.6  & 14.14  & 2.513 & 100 & No & No & No \\

13 & 1000 & 1.0 & $2\pi$    & 2.6  & 14.14  & 12.57 & 100 & No & No & No \\

14 & 1000 & 1.0 & $\pi$/5 & 2.6  & 14.14   & 1.256 & 25& No & No & No  \\

15 & 1000 & 1.0 & $\pi$/5 & 2.6  & 14.14  &1.256 & 50& No & Yes & Yes  \\

16 & 1000 & 1.0 & $\pi$/5 & 2.6  & 14.14   & 1.256 & 90& Yes & Yes & Yes \\

17 & 1000 & 1.0 & $\pi$/5 & 2.6 & 14.14   & 1.256 & 150& Yes & Yes & Yes  \\

\end{tabular}
\end{ruledtabular}
}
\end{table}

\nocite{*}

\bibliography{apssamp}

\end{document}